\documentclass[10pt,preprint2,longnamesfirst]{aastex}
\received{2007 February~1}
\revised{2007}
\accepted{2007 June~1}
\paperid{71366}

\newcommand{\y}{Y${}_1$}

\shorttitle{}
\shortauthors{Lazio et al.}

\begin{document}

\title{Angular Broadening of Intraday Variable \hbox{AGN}.~\hbox{II}.
Interstellar and Intergalactic Scattering}
\author{T.~Joseph~W.~Lazio}
\affil{Naval Research Laboratory, 4555 Overlook Ave.~\hbox{SW},
	Washington, DC 20375-5351; USA} 
\email{Joseph.Lazio@nrl.navy.mil}

\author{Roopesh Ojha}
\affil{NVI/United States Naval Observatory, 3450 Massachusetts 
Ave.~\hbox{NW}, Washington DC 20392-5420; USA}
\email{rojha@usno.navy.mil}

\author{Alan L.~Fey}
\affil{United States Naval Observatory, 3450 Massachusetts
	Ave.~\hbox{NW}, Washington DC 20392-5420; USA}
\email{afey@usno.navy.mil}

\author{Lucyna Kedziora-Chudczer}
\affil{Institute of Astronomy, School of Physics A28, The 
University of Sydney, NSW 2006; Australia}

\author{James M.~Cordes}
\affil{Department of Astronomy, Cornell University and \hbox{NAIC},
	Ithaca, NY 14853; USA}
\email{cordes@astro.cornell.edu}

\and

\author{David L.~Jauncey and James~E.~J.~Lovell}
\affil{Australia Telescope National Facility \hbox{CSIRO}, PO~Box~76, Epping, NSW 1710; Australia}

\begin{abstract}
We analyze a sample of~58 multi-wavelength, Very Long Baseline Array
observations of active galactic nuclei (AGN) to determine their
scattering properties.  Approximately 75\% of the sample consists of
AGN that exhibit centimeter-wavelength intraday variability
(interstellar scintillation) while the other 25\% do not show intraday
variability.  We find that interstellar scattering is measurable for
most of these \hbox{AGN}, and the typical broadening diameter is~2~mas
at~1~GHz.  We find that the scintillating AGN are typically at lower
Galactic latitudes than the non-scintillating \hbox{AGN}, consistent
with the scenario that intraday variability is a propagation effect
from the Galactic interstellar medium.  The magnitude of the inferred
interstellar broadening measured toward the scintillating \hbox{AGN},
when scaled to higher frequencies, is comparable to the diameters
inferred from analyses of the light curves for the more well-known
intraday variable sources.  However, we find no difference in the
amount of scattering measured toward the scintillating versus
non-scintillating \hbox{AGN}.  A consistent picture is one in which
the scintillation results from localized regions (``clumps'')
distributed throughout the Galactic disk, but which individually make
little contribution to the angular broadening.  Of the 58 AGN
observed, 37 (64\%) have measured redshifts.  At best, a marginal
trend is found for scintillating (non-scintillating) AGN to have
smaller (larger) angular diameters at higher redshifts.  We also use
our observations to try to constrain the possibility of intergalactic
scattering.  While broadly consistent with the scenario of a highly
turbulent intergalactic medium, our observations do not place
significant constraints on its properties.
\end{abstract}


\section{Introduction}\label{sec:intro}

There is now compelling evidence that the intraday variability (IDV)
phenomenon---intensity variations on hour time scales at centimeter
wavelengths in compact, flat-spectrum, extragalactic sources
\citep[e.g.,][]{h84,qwk+92}---is of extrinsic origin.  
Density fluctuations in the interstellar medium (ISM) induce
refractive index fluctuations which, when combined with the relative
motions of the scattering medium and the Earth, produce intensity
variations or scintillations.  The evidence for this conclusion is
both differences in the variability pattern arrival times at widely
spaced radio telescopes and annual cycles in the variability
characteristics for various IDV sources
\citep{jblk-ctmr00,jm01,rwkkq01,d-td02,d-td03,jjblk-ctm03,jlkk05,bmjltk-c06}.

In order to exhibit interstellar scintillations (ISS), a source must
contain a sufficiently compact component (analogous to ``Stars
twinkle, planets don't'') such that its angular diameter is comparable with
or smaller than the size of the first Fresnel zone of the scattering
screen, i.e., of order tens of microarcseconds at frequencies near a
few gigahertz \citep[e.g.,][]{w98}.  The \emph{absence} of ISS in a
source could be either an intrinsic or extrinsic effect.  Active
galactic nuclei (AGN) might have a range of intrinsic diameters, in
which case only the most compact would exhibit \hbox{ISS}.
Alternately, interstellar density fluctuations produce a rich range of
observable phenomena \citep{r90}, of which scintillations are only one
manifestation.  Angular broadening along the line of sight, due either
to multiple ionized media or an extended medium, could produce
apparent diameters of AGN sufficiently large that the AGN would not
display \hbox{ISS}.

Consistent with this requirement for a compact ($\sim 10$~$\mu$as) component, 
\cite{ofjlj04} have compared AGN that display ISS with those that
do not and find that the scintillating AGN typically are more
core-dominated on a milliarcsecond scale than the non-scintillating
\hbox{AGN}. This result is striking given that their observations
compared the source structure on milliarcsecond, not microarcsecond,
scales.  However, the observations of \cite{ofjlj04} were at the
single frequency of~8.4~GHz.  This frequency is sufficiently high that
interstellar scattering effects on most lines of sight through the ISM
would not be detectable on terrestrial baselines nor could they have
exploited the wavelength dependence for scattering to separate the its
effects from the intrinsic diameters of the sources.

The AGN observed by \cite{ofjlj04} were drawn from the
Micro-Arcsecond Scintillation-Induced Variability (MASIV) survey
\citep{ljbk-cmrt03}.  The MASIV survey is a large variability survey
of the northern sky with the primary goal being the construction of a
large sample of scintillating \hbox{AGN}.  The survey used the Very
Large Array (VLA) at~5~GHz in a multi-array mode and has yielded
scintillation information on over~500 \hbox{AGN}, of which over half
have been found to be scintillating \citep{ljssbmrk-c07}.

\cite{ofljlk-c06} presented multi-frequency
observations of a subset of MASIV sources.  Their observations were
designed to create a sample of sufficient size to compare and contrast
the scattering behavior of scintillating and non-scintillating
\hbox{AGN}.  This paper reports the analysis of those observations.

\section{Source Sample}\label{sec:observe}

Our sample consists of~58 \hbox{AGN}, observed in two different
programs.  The first subset consists of~49 AGN from the MASIV survey,
observed over~3~days with the Very Long Baseline Array (VLBA) in~2003
February.  At the time of the observations, approximately half of
these 49 AGN were classified as highly variable MASIV sources, with
scintillation indices larger than 2\%; the other half were classified
as non-scintillators, with no scintillation index larger than 0.2\%.
These sources were chosen without regard to their Galactic latitude or
longitude.  Since our observations, however, further MASIV
observations and analysis shows that many of the AGN identified
originally as non-scintillating are in fact scintillating
\citep{ljssbmrk-c07}.  The number of recognized scintillating AGN is
now 46, and the number of non-scintillating sources is now 13 (ratio
of~3:1).  The subset of AGN
from the MASIV survey were observed at~0.33, 0.61, and~1.6~GHz, with
additional observations at~2.3 and~8.4~GHz.  In total, 194 VLBA images
of the 49 MASIV extragalactic radio sources at up to 7 observing
frequencies were obtained \citep{ofljlk-c06}.  Additional data were
obtained from the United States Naval Observatory (USNO) Radio
Reference Frame Image Database\footnote{
\tt{http://www.usno.navy.mil/RRFID/}}
(RRFID) and the literature.

The observations of these 49 AGN were acquired by cycling through the
sources so as to increase the $u$-$v$ plane coverage.  Typical times
on source range from~10~min. at the higher frequencies to~25~min. at
the lower frequencies.  Typical noise levels were within a factor
of~2--3 of the expected thermal noise limits.

\cite{ofljlk-c06} fit gaussian component models to the visibility data
of the sources, using the images as guides.  If more than one
component was required to model a source at a given frequency, the
most compact component was identified as the ``core,'' as the most
compact component will be the one for which scattering effects will be
most apparent.  The most compact component is frequently the brightest
one.  For the few sources where this does not hold strictly at all
frequencies, the compact and bright component that could consistently
be identfied as the same at all frequencies was identified as the
``core,'' e.g., \objectname[]{J0713$+$4349} where the northernmost
component is identified as the core even though it is not the
brightest component at~8.4~GHz \citep{ofljlk-c06}.

The second subset consists of~9 AGN chosen from those used to define
the International Celestial Reference Frame
\citep{johnstonetal95,maetal98}.  The initial motivation was to use
these AGN to search for scattering resulting from the interstellar
media of galaxies along their lines of sight.  The sources were chosen
to be (1)~At Galactic latitudes $|b| > 45\arcdeg$; (2)~Strong, with
$S_{\mathrm{6\,cm}} \approx 1$~Jy; and~(3)~Compact and dominated by a
single component \citep{fc97}.  Observations of these 9 sources were
conducted with the VLBA on~2001 February~17 and~18 at~0.33
and~0.61~GHz.  Calibration, imaging, and model extraction was
performed in a manner similar to that used by \cite{ofljlk-c06}.

Because both subsets involved observations at~0.33 and~0.61~GHz, the
observations were generally carried out at night so that the sources
had large solar elongations.  Indeed, observing at large solar
elongation was an explicit criterion in scheduling the observations
for the second subset.  As a result, the smallest elongation for any
source is 75\arcdeg, and the typical elongation is approximately
130\arcdeg.

For the present analysis, we have used the core components models from
these two observing programs, augmented by measurements from the
literature.  All sources have angular diameters measured at at least 3
frequencies, and some sources have measured angular diameters at as
many as 7 frequencies.  See Table~2 of \cite{ofljlk-c06}.  While the
selection criteria for the two subsets differed, the sources were
treated identically in the following analysis.

\section{Analysis}\label{sec:analyze}

Angular broadening is manifested as an observed angular diameter that
scales approximately as $\lambda^2$.  We fit the measured angular
diameters to the functional form
\begin{equation}
\theta^2 = (\theta_s\nu^{-2.2})^2 + (\theta_i\nu^x)^2
\label{eqn:fit}
\end{equation}
where $\theta_s$ and~$\theta_i$ are the scattering and intrinsic
(FWHM) diameters of the \hbox{AGN}, respectively, at the fiducial
frequency of~1~GHz.  We found the best-fitting values for~$\theta_s$
and~$\theta_i$ in a minimum $\chi^2$ sense.  We considered both $x =
0$ (i.e., frequency-independent intrinsic diameter, for a flat
spectrum source) and $x = -1$ (i.e., frequency scaling for a single
incoherent synchrotron component) and selected the value of~$x$ that
produced the lower $\chi^2$.

As a motivation for the use of equation~(\ref{eqn:fit}), as well as
anticipating later discussion, we begin by considering a crude
approximation to equation~(\ref{eqn:fit}).  For the sources having
measured angular diameters at both 0.33 and~1.6~GHz, we have assumed
a simple power-law scaling for the angular diameter, $\theta \propto
\nu^\beta$, and solved for~$\beta$.  We chose 0.33~GHz because the
frequency dependence of scattering means that it will be the most
sensitive to scattering; we chose 1.6~GHz as the second frequency as
an attempt to balance between having a sufficiently large frequency
dynamic range so as to obtain a robust estimate of~$\beta$ but not
having such a large frequency range that intrinsic structure might
dominate.  If scattering is important, we expect $\beta \approx -2$.
We find an average value of $\bar\beta = -1.9 \pm 0.1$ for the entire
sample.  Clearly, we anticipate that intrinsic structure may be
important for some sources, but that the $\bar\beta$ is close to the
expected value from scattering indicates that scattering is important
for the sample of sources as a whole.

\begin{figure}[tb]
\epsscale{0.55}
\begin{center}
\rotatebox{-90}{\plotone{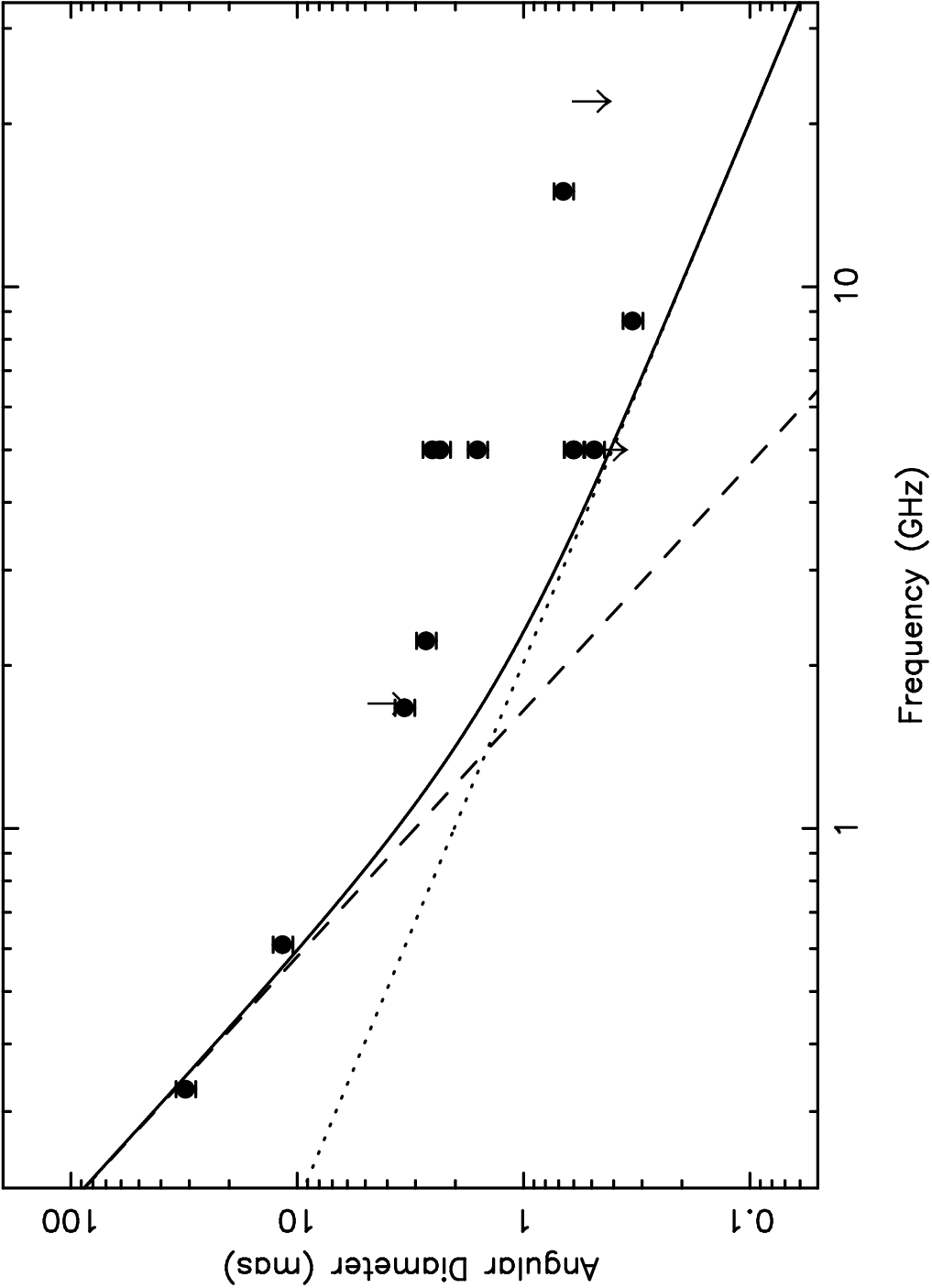}}
\rotatebox{-90}{\plotone{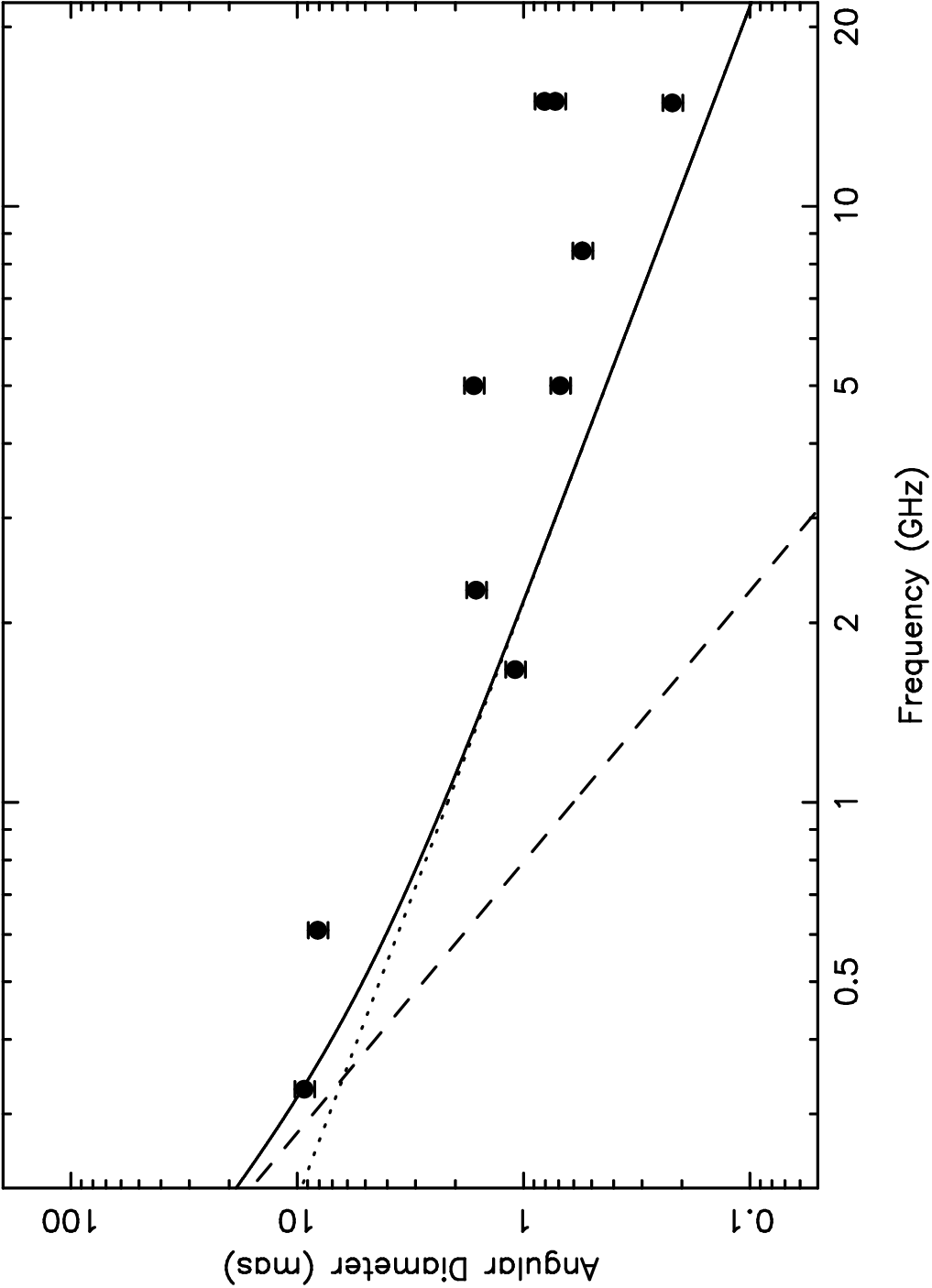}}
\end{center}
\vspace*{-3ex}
\caption[]{Two examples of fitting results.  For both plots, solid
circles show measured diameters, arrows indicate upper limits, the
solid line indicates the fit of equation~(\ref{eqn:fit}) to the
observations, the dashed line indicates the inferred scattering
diameter, and dotted line indicates the inferred intrinsic diameter.
Uncertainties on the angular diameters are plotted, but in many cases
are comparable to the size of the symbol.
\textit{Top} The source \objectname[]{J2022$+$6136} for which a
relatively large scattering diameter, 3.1~mas at~1~GHz, is inferred.
\textit{Bottom} The source \objectname[]{J0745$+$1011} for which a
relatively small scattering diameter, 0.6~mas at~1~GHz, is inferred.}
\label{fig:example}
\end{figure}

Table~\ref{tab:fit} summarizes the inferred scattering and intrinsic
diameters from fitting the data for each source to
equation~(\ref{eqn:fit}).  For comparison, we also list the predicted
diameter at~1~GHz from the NE2001 model \citep{cl02} for interstellar
scattering, $\theta_{\mathrm{NE2001}}$.  Figure~\ref{fig:example}
illustrates examples of the measured angular diameters, showing the
results for an AGN for which a relatively large amount of scattering
is inferred and one for which a relatively small amount of scattering
is inferred.

\begin{deluxetable}{lcccccccccc}
\tablecaption{Fitted Scattering and Intrinsic
	Diameters\label{tab:fit}}
\tablewidth{0pc}
\tabletypesize{\small}
\tablehead{%
 \colhead{Name} & \colhead{$\ell$}      & \colhead{$b$}
	& \colhead{$z$} & \colhead{Scintillate?} & \colhead{$x$} & \colhead{$\theta_s$} &
	\colhead{$\theta_i$} & \colhead{$N$} & \colhead{$\chi^2$} & 
	\colhead{$\theta_{\mathrm{NE2001}}$} \\
                & \colhead{({}\arcdeg)} & \colhead{({}\arcdeg)}
        &               &                        &               & \colhead{(mas)}      &
	\colhead{(mas)}      &               & 	                  &  
	\colhead{(mas)}}
\startdata

\objectname[MASIV]{J0102$+$5824} & 124.419 & $-4.435$  & 0.644   & Y  & 1 & 3.6 & 1.2 &  6 &  6 &  4.9 \\
\objectname[MASIV]{J0217$+$7349} & 128.927 &  11.964   & 2.367   & N  & 1 & 2.8 & 0.5 &  9 & 11 &  2.0 \\
\objectname[MASIV]{J0343$+$3622} & 157.530 & $-14.691$ & 1.484   & Y  & 1 & 3.2 & 3.7 &  5 & 39 &  1.9 \\
\objectname[MASIV]{J0349$+$4609} & 152.152 &  $-6.369$ & \nodata & N  & 1 & 2.1 & 1.2 &  4 &  5 &  4.1 \\
\objectname[MASIV]{J0403$+$2600} & 168.025 & $-19.648$ & 2.109   & \y & 0 & 2.4 & 0.3 &  8 & 17 &  1.5 \\
				 				          
\\				 				          
				 
\objectname[MASIV]{J0419$+$3955} & 160.461 &  $-7.336$ & \nodata & Y  & 0 & 2.6 & 0.2 &  3 &  2 &  3.8 \\
\objectname[MASIV]{J0423$+$4150} & 159.705 &  $-5.381$ & 2.277   & Y  & 0 & 4.1 & 0.5 &  3 & 0.3 & 4.6 \\
\objectname[MASIV]{J0451$+$5935} & 149.323 &   9.660   & \nodata & Y  & 0 & 4.9 & 0.2 &  3 &  8 &  2.9 \\
\objectname[MASIV]{J0502$+$1338} & 187.414 & $-16.745$ & \nodata & Y  & 0 & 3.7 & 0.1 &  3 &  1 &  1.6 \\
\objectname[MASIV]{J0503$+$0203} & 197.911 & $-22.815$ & \nodata & N  & 0 & 4.6 & 0.5 &  8 & 10 &  1.2 \\
				        			          
\\				        			          
				        			          
\objectname[MASIV]{J0507$+$4645} & 161.025 &   3.716   & \nodata & \y & 0 & 1.6 & 1.0 &  3 &  9 &  5.2 \\
\objectname[MASIV]{J0509$+$0541} & 195.405 & $-19.635$ & \nodata & Y  & 0 & 3.0 & 0.2 &  5 & 24 &  1.3 \\
\objectname[MASIV]{J0539$+$1433}\tablenotemark{a} & 191.597 & $-8.660$ & 2.69 & Y & 1 & 3.4 & 2.4 & 3 & 3 & 3.1 \\
\objectname[MASIV]{J0539$+$1433}\tablenotemark{a} & 191.597 & $-8.660$ & 2.69 & Y & 0 & 6.1 & 0.4 & 3 & 3 & 3.1 \\
\objectname[MASIV]{J0607$+$6720} & 146.804 &  20.858   & 1.97    & \y & 0 & 1.4 & 0.4 & 11 & 20 &  1.3 \\
				   
\\				   
				   
\objectname[MASIV]{J0650$+$6001} & 155.842 &  23.155   & 0.455   & Y  & 0 & 7.8 & 0.3 &  6 & 36 &  1.3 \\
\objectname[MASIV]{J0654$+$5042} & 165.680 &  21.106   & \nodata & \y & 0 & 1.8 & 0.2 &  3 & 45 &  1.4 \\
\objectname[MASIV]{J0713$+$4349} & 173.792 &  22.199   & 0.518   & N  & 1 & 1.3 & 4.7 & 11 & 18 &  1.4 \\
\objectname[MASIV]{J0721$+$7120} & 143.981 &  28.017   & 2.06    & \y & 1 & 0.6 & 0.9 &  8 & 39 &  1.1 \\
\objectname[MASIV]{J0725$+$1425} & 203.643 &  13.908   & \nodata & Y  & 0 & 1.5 & 0.2 &  6 & 36 &  1.7 \\
				        			          
\\				       			          
				        			          
\objectname[MASIV]{J0738$+$1742} & 201.846 &  18.070   & 0.424   & \y & 0 & 1.4 & 0.2 &  9 & 38 &  1.5 \\
\objectname[MASIV]{J0745$+$1011} & 209.796 &  16.592   & 2.624   & \y & 1 & 0.6 & 2.2 & 10 & 36 &  1.4 \\
\objectname[MASIV]{J0757$+$0956} & 211.311 &  19.057   & 0.266   & Y  & 0 & 1.1 & 0.2 &  9 & 24 &  1.3 \\
\objectname[MASIV]{J0808$+$4950} & 169.163 &  32.564   & 1.418   & \y & 1 & 1.1 & 1.1 & 12 & 35 &  1.1 \\
\objectname[MASIV]{J0830$+$2410} & 200.021 &  31.876   & 0.939   & Y  & 1 & 0.8 & 1.9 &  6 &  9 &  1.2 \\
				        			          
\\				        			          
				        			          
\objectname[MASIV]{J0831$+$0429} & 220.693 &  24.331   & 0.180   & \y & 1 & 4.8 & 1.0 &  9 & 18 &  1.1 \\
\objectname[MASIV]{J0842$+$1835} & 207.275 &  32.480   & 1.270   & \y & 0 & 4.1 & 0.5 &  8 & 23 &  1.1 \\
\objectname[MASIV]{J0914$+$0245} & 228.352 &  32.819   & 0.427   & Y  & 0 & 2.8 & 0.1 &  4 & 16 &  1.0 \\
\objectname[MASIV]{J0920$+$4441} & 175.700 &  44.815   & 2.190   & N  & 1 & 1.7 & 0.7 & 10 & 36 &  1.0 \\
\objectname[MASIV]{J0956$+$2515} & 205.511 &  50.981   & 0.712   & Y  & 1 & 0.0 & 2.3 & 11 & 31 &  0.9 \\
				        			          
\\				        			          
				        			          
\objectname[MASIV]{J0958$+$4725} & 170.055 &  50.730   & 1.882   & Y  & 0 & 2.3 & 0.2 &  7 &  7 &  0.9 \\
\objectname[MASIV]{J1008$+$0621} & 233.521 &  46.012   & \nodata & Y  & 0 & 5.0 & 0.0 &  3 &  8 &  0.9 \\
\objectname[MASIV]{J1014$+$2301} & 210.699 &  54.431   & 0.565   & Y  & 0 & 3.5 & 0.3 &  7 & 37 &  0.9 \\
\objectname[MASIV]{J1041$+$5233} & 157.521 &  54.965   & 0.677   & Y  & 0 & 1.4 & 0.8 &  4 &  1 &  0.8 \\
\objectname[MASIV]{J1125$+$2610} & 210.920 &  70.885   & 2.341   & \y & 1 & 0.6 & 1.1 &  5 & 18 &  0.6 \\
				        			          
\\				        			          
				        			          
\objectname[MASIV]{J1153$+$8058} & 125.719 &  35.836   & 1.250   & \y & 0 & 1.9 & 0.2 &  6 & 17 &  0.9 \\
\objectname[MASIV]{J1159$+$2914} & 199.413 &  78.374   & 0.729   & Y  & 1 & 2.7 & 1.5 &  8 & 18 &  0.5 \\
\objectname[MASIV]{J1327$+$2210} &   3.380 &  80.527   & 1.400   & N  & 1 & 1.6 & 1.2 &  4 & 28 &  0.6 \\
\objectname[MASIV]{J1407$+$2827} &  41.862 &  73.251   & 0.076   & N  & 1 & 0.0 & 5.0 &  8 &  9 &  0.5 \\
\objectname[MASIV]{J1642$+$6856} & 100.705 &  36.621   & 0.751   & Y  & 1 & 1.2 & 1.7 & 13 & 46 &  0.8 \\
				        			          
\\				       			          
				        			          
\objectname[MASIV]{J1656$+$6012} &  89.627 &  37.430   & 0.623   & Y  & 0 & 6.8 & 0.3 &  3 & 0.1 & 0.8 \\
\objectname[MASIV]{J1746$+$6226} &  91.621 &  31.320   & 3.889   & \y & 0 & 2.7 & 0.2 &  5 & 19 &  0.8 \\
\objectname[MASIV]{J1812$+$5603} &  84.587 &  27.473   & \nodata & Y  & 0 & 3.3 & 0.3 &  3 & 12 &  0.8 \\
\objectname[MASIV]{J1823$+$6857} &  99.210 &  27.669   & \nodata & Y  & 1 & 1.8 & 1.6 &  4 & 12 &  0.9 \\
\objectname[MASIV]{J1927$+$6117} &  92.726 &  19.446   & \nodata & \y & 1 & 1.8 & 1.5 &  5 &  1 &  1.0 \\
 				 				          
\\				 				          
				 				          
\objectname[MASIV]{J2002$+$4725} &  82.219 &   8.793   & \nodata & \y & 1 & 0.0 & 2.7 &  3 &  4 &  1.8 \\
\objectname[MASIV]{J2009$+$7229} & 105.355 &  20.180   & \nodata & Y  & 0 & 1.3 & 0.3 &  3 & 21 &  1.1 \\
\objectname[MASIV]{J2022$+$6136} &  96.082 &  13.775   & 0.227   & N  & 0 & 3.1 & 0.4 & 15 & 30 &  1.4 \\
\objectname[MASIV]{J2230$+$6946} & 111.248 &  10.164   & \nodata & Y  & 0 & 2.4 & 0.3 &  7 & 16 &  2.0 \\
\objectname[MASIV]{J2311$+$4543} & 105.315 & $-13.703$ & 1.447   & Y  & 0 & 2.4 & 0.4 &  5 & 26 &  1.4 \\
				 				          
\\				 				          
				 				          
\objectname[RORF]{B0955$+$476} 	 & 170.055 &   50.730  & 1.873   & Y  & 1 & 2.1  & 0.1  &  7 & 34  & 1.0 \\
\objectname[RORF]{B1130$+$009} 	 & 264.364 &   57.582  & \nodata & N  & 0 & 1.8  & 0.9  &  4 & 51  & 0.9 \\
\objectname[RORF]{B1226$+$373} 	 & 147.142 &   78.938  & 1.515   & N  & 1 & 2.8  & 0.1  &  4 & 15  & 0.5 \\
\objectname[RORF]{B1236$+$077} 	 & 294.112 &   70.170  & 0.400   & N  & 1 & 1.4  & 0.3  &  3 & 48  & 0.8 \\ 
\objectname[RORF]{B1402$+$044} 	 & 343.669 &   61.169  & 3.211   & N  & 1 & 2.2  & 0.3  &  6 & 48  & 0.6 \\
				  				          
\\				  				          
				  				          
\objectname[RORF]{B1432$+$200}   &  21.387 &   65.299  & \nodata & \y & \nodata & 2.3 & 0.0 & 3 & \nodata & 0.5 \\
\objectname[RORF]{B1459$+$480}   &  81.122 &   57.419  & \nodata & Y  & 0       & 1.6 & 1.1 & 4 & 110     & 0.8 \\
\objectname[RORF]{B1502$+$036}   &   2.226 &   50.254  & 0.413   & \y & \nodata & 1.0 & 2.9 & 3 & 0.2     & 0.7 \\
\objectname[RORF]{B1502$+$106}   &  11.381 &   54.580  & 1.833   & N  & 1       & 2.4 & 0.2 & 6 & 12      & 0.6 \\
			      
\enddata		      
\tablenotetext{a}{Both $\nu^{-1}$ and $\nu^0$ dependences for the
intrinsic diameter produced equal $\chi^2$.}
\tablecomments{Sources are indicated to be either scintillators (Y),
non-scintillators (N), or sources observed to vary once and presumed
to be scintillators (\y); $x$ is the spectral index for the frequency
dependence of the intrinsic diameter, equation~(\ref{eqn:fit});
$\theta_s$ and~$\theta_i$ are the inferred scattering and intrinsic
diameters, respectively; $N$ and $\chi^2$ are the number of data and
chi-square in the fit for equation~(\ref{eqn:fit}); and
$\theta_{\mathrm{NE2001}}$ is the predicted interstellar scattering
diameter from the NE2001 model.}
\end{deluxetable}

There are values of the reduced $\chi^2$ in Table~\ref{tab:fit} that
are larger than unity, at times by a significant factor.  These result
from a combination of two factors.  First, the uncertainties for some
data are likely to be underestimated.  For the observed diameters
obtained from our observations, we estimated their uncertainties using
a bootstrap-like procedure in which visibility data associated with
different antennas were removed before performing the fit.  The range
of fitted diameters suggests a 10\% uncertainty.  For angular
diameters obtained from the literature, we have assumed the same
fractional uncertainty.  However, we have been able to identify data
for which this assumed 10\% is likely to be too small.  

The fitting procedure also yields an estimate of the uncertainty in
the inferred scattering diameter (at~1~GHz).  The median value of this
uncertainty is 0.1~mas.  We have repeated the fitting procedure with
larger uncertainties adopted for the angular diameters obtained from
the literature, and in some cases even removing apparent outliers.
The typical change in the inferred diameter is comparable to the
uncertainty in the inferred scattering diameter.  We have also
repeated analyses described below with larger uncertainties for the
angular diameters from the literature and find no change from the
results we present below.

A second potential cause of large $\chi^2$ values is that we have
adopted fixed frequency scaling exponents in equation~(\ref{eqn:fit}).
We might obtain a better fit by letting $x$ be a fitted parameter,
fitting the scattering frequency dependence rather than adopting
$-2.2$, or both.  To do so would often require more data than are
available.  Consequently, while larger than unity, we consider there
to be plausible explanations for the $\chi^2$ values in
Table~\ref{tab:fit} and shall use the angular diameters resulting from
our fits.

\cite{ofljlk-c06} divided the sources into two groups,
``scintillators'' and ``non-scintillators.''  Since the publication of
that paper, it has been realized that some of the AGN identified as
scintillators displayed variation at only a single epoch (``once-er'')
among the MASIV observations, leading to the possibility that a
non-scintillator would be classified mistakenly as a scintillator.
Analysis of the MASIV observations (to be published elsewhere)
suggests that the light curve from an individual epoch can be
classified correctly at the 95\% confidence level.  From the four
epochs of observations comprising \hbox{MASIV}, the probability of a
false identification is only 4.3\%, meaning that we expect only 1
source in our sample to be classified mistakenly.  While we identify
the single-epoch variable sources in Table~\ref{tab:fit}, their
presence should have have no significant effect on our analysis, and
we treat the single-epoch variable AGN as scintillators.  Also,
analysis has continued on the MASIV sources, so there may be
occasional differences in the classification (scintillating vs.\
non-scintillating) in our Table~\ref{tab:fit} as compared to Table~1
of \cite{ofljlk-c06}.

\begin{figure}
\epsscale{0.5}
\rotatebox{-90}{\plotone{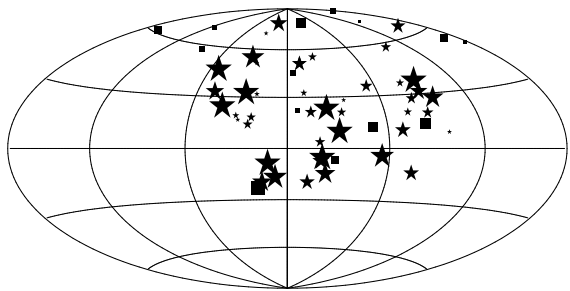}}
\caption[]{The distribution of sources observed as a function of
Galactic coordinates.  The Galactic \emph{anticenter} is at the center of the
plot, and longitude increases to the left.  Stars indicate sources
that scintillate; squares indicate non-scintillating sources.  The
size of the symbol is qualitatively proportional to the inferred scattering diameter.}
\label{fig:distribute}
\end{figure}

Figure~\ref{fig:distribute} shows the distribution of the inferred
scattering diameters plotted as a function of Galactic coordinates.
We have considered the distribution of the sources on the sky as a
function of both Galactic latitude~$b$ and ecliptic latitude~$\beta$
and the distribution as a function of solar elongation.
There is no statistically significant correlation of the inferred
scattering diameter with either coordinate (strictly, the absolute
value of the coordinate), nor with $\cos(|b|)$ or $\cos(|\beta|)$.
(Typical correlation coefficients are approximately 0.1.)  The use of
the cosine of the angle [$\cos(|b|)$ or $\cos(|\beta|)$] attempts to
compensate for the increased amount of sky at low latitudes.  There is
also no correlation between inferred scattering diameter and solar elongation.

The lack of a correlation of the scattering diameter with Galactic
latitude would appear to be at odds with previous determinations that
interstellar scattering increases rapidly at low latitudes
\citep[e.g.,][]{ra84}.  However, while no longitude selection
criterion was applied in constructing our sample, our sample is
nonetheless weighted strongly toward the outer Galaxy.  We have no
sources with longitudes in the range $-60\arcdeg < \ell < 60\arcdeg$,
and only a few at low latitudes in the range $-120\arcdeg < \ell <
120\arcdeg$.  Thus, we attribute the apparent lack of a correlation
between scattering diameter and Galactic latitude as a result of
having few, essentially no, lines of sight into the inner Galaxy.

While there is no correlation of scattering diameter with Galactic
latitude for the entire set of sources, scintillating AGN are
consistently at lower Galactic latitudes than the non-scintillating
\hbox{AGN}.  Table~\ref{tab:stats} shows that the average absolute values
for the Galactic latitudes of scintillating and non-scintillating
sources differ by nearly 20\arcdeg.  No such difference is found for
the average (absolute) ecliptic latitude.

\begin{deluxetable}{lcc}
\tablecaption{Statistical Measures\label{tab:stats}}
\tablewidth{0pc}
\tablehead{ & \colhead{Scintillators} & \colhead{Non-scintillators}}
\startdata
             & \multicolumn{2}{c}{Galactic latitude (absolute value)} \\
mean         & $29\fdg5 \pm 3\fdg0$ & $46\fdg3 \pm 7\fdg0$ \\
\\
             & \multicolumn{2}{c}{ecliptic latitude (absolute value)} \\
mean         & $34\fdg9 \pm 4\fdg3$ & $28\fdg8 \pm 4\fdg9$ \\
\\
             & \multicolumn{2}{c}{Scattering Diameters (at~1~GHz)} \\
mean (mas)   & 2.1 $\pm$ 0.2 & 2.1 $\pm$ 0.2 \\
median (mas) & 1.9           & 2.0 \\
\\
             & \multicolumn{2}{c}{Redshifts} \\
mean (mas)   & 1.38 $\pm$ 0.18 & 1.29 $\pm$ 0.31 \\
median (mas) & 1.27            & 1.40 \\

\enddata
\end{deluxetable}

\begin{figure}
\epsscale{0.7}
\begin{center}
\rotatebox{-90}{\plotone{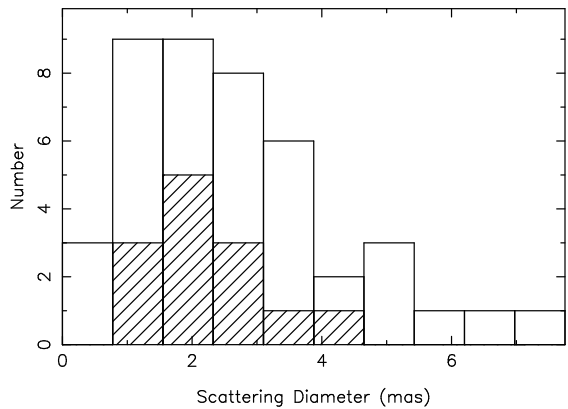}}
\end{center}
\vspace*{-3ex}
\caption[]{The histograms of scattering diameters.  The open histogram
shows the distribution for the scintillating sources; the hatched
histogram shows the distribution for the non-scintillating sources.}
\label{fig:histo}
\end{figure}

Figure~\ref{fig:histo} shows a histogram of the inferred scattering
diameters.  We have used a Kolmogorov-Smirnov test to assess whether
scintillating AGN have a different distribution of inferred scattering
diameters as compared to the non-scintillating \hbox{AGN}.  We find no
statistical difference: the scattering diameters of scintillating AGN
do not differ appreciably from those of the non-scintillating
\hbox{AGN}.

Examination of the scattering diameters inferred for individual
objects indicates that some of the largest scattering diameters result
from AGN for which no measurements exist below~1~GHz.  In order
that these not bias our result, we removed these and repeated the K-S
test analysis.  There is no change in the result, that the scattering
diameters for the scintillating and non-scintillating AGN are
consistent with having been drawn from the same distribution.
Both the mean and the median scattering diameter for scintillating and
non-scintillating sources is approximately 2~mas (Table~\ref{tab:stats}).

\begin{figure}[tb]
\epsscale{0.7}
\rotatebox{-90}{\plotone{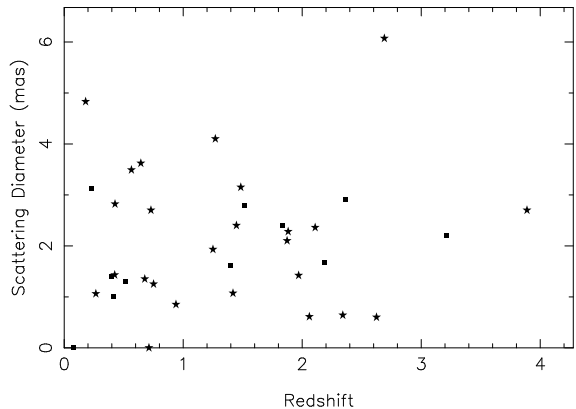}}
\vspace*{-1ex}
\caption[]{The distribution of scattering diameters as a function of
redshift.  Stars indicate sources that scintillate; squares indicate
non-scintillating sources.}
\label{fig:z}
\end{figure}

From the entire sample, 37 AGN (64\%) have measured redshifts.  There
appears to be little difference in the redshift distribution of the
scintillating and non-scintillating \hbox{AGN}, with the two
populations having similar means and medians (Table~\ref{tab:stats}).
Figure~\ref{fig:z} shows the distribution of the scattering diameters
as a function of redshift.  We have determined the correlation between
the scattering diameters and redshifts for the entire sample, as well
as splitting it into the two populations, scintillating and
non-scintillating.  There is no correlation of the scattering diameter
with redshift for the entire sample.  There may be a marginal
correlation, at the 5\% confidence level, between the scattering
diameters and redshift, in the opposite sense for the scintillating
and non-scintillating sources.  That is, the scattering diameters of
scintillating (non-scintillating) AGN may become smaller (larger) at
higher redshifts.

\section{Discussion and Conclusions}\label{sec:conclude}

In one sense, our results are broadly consistent with what is known
about intraday variability and interstellar scattering.  In our
sample, scintillating AGN lie typically at lower Galactic latitudes,
consistent with the notion that the scintillation responsible for IDV
results from the Galactic ISM (\S\ref{sec:intro}).

In addition to the \hbox{ISM}, other ionized media along the lines of
sight to these sources are the interplanetary medium (IPM) and the
intergalactic medium (IGM).  It is possible that the IPM could
contribute angular broadening at a level sufficient to be detectable
in this analysis, particularly given the use of~0.33 and~0.61~GHz
observations.  The lack of a correlation with either the ecliptic
latitude (Table~\ref{tab:stats}) or solar elongation, however, implies
that the IPM makes no detectable contribution in these observations.
The lack of a strong correlation with redshift and the systematically
lower Galactic latitude of the broadened sources suggests that the IGM
makes no detectable contribution to our observations, either.

In a further effort to differentiate the Galactic contribution of
scattering from any possible intergalactic contribution, we have
searched for pulsars within~1\arcdeg\ of the AGN in our sample.  We
find no pulsars this close to any of our sources.  Given the
relatively low density of pulsars on the sky, a significantly larger
sample of AGN would be required in order to make such a comparison.

On the face of it, the result that scintillating AGN are broadened at
levels comparable to those of the non-scintillating AGN appears to
contradict the requirement (\S\ref{sec:intro}) that in order to
display scintillation, a source must be sufficiently compact.  One
possibility is that scattering is, in fact, not important at all and
that we have mistakenly attributed the effects of intrinsic structure
to angular broadening.  We regard this as unlikely, given that the
angular diameters of the sample of sources, as a whole, scale
approximately as expected from interstellar scattering.

In fact, the magnitude of the estimated broadening does not appear to
be problematic from the standpoint of quenching the scintillations.
Taking 2~mas at~1~GHz as a characteristic scattering diameter
(Table~\ref{tab:stats}), the implied scattering diameter at~5~GHz is
80~$\mu$as, comparable to the value derived for a number of the
well-known scintillating sources
\citep[e.g.,][]{rqwkw95,rk-cj02,d-td03,bjl+03}.

A notable feature of the most extreme IDV sources is that detailed
analyses of their light curves suggest that the scattering medium
responsible for the scintillation lies quite close to the Sun
($\approx 25$~pc).  If local material was responsible for the
scintillation of all scintillating \hbox{AGN}, we would not expect a
difference between the average Galactic latitude of scintillating and
non-scintillating sources (Table~\ref{tab:stats}).  That such a
difference exists suggests that the scintillation for most
scintillating AGN results from scattering material associated with the
Galactic \hbox{ISM}. 

A consistent explanation for these results is obtained if the
scintillation is produced from small ``clumps'' of scattering
material, distributed throughout the Galactic disk.  For instance,
\cite{d-td03} determine that a ``thin screen'' is responsible for the
extreme IDV of \objectname[]{J1819$+$3845}, with the screen being
about~10~pc distant and having an internal level of scattering
measured by $C_n^2 = 0.5$~m${}^{-20/3}$.  They do not provide a
quantitative estimate of the thickness of this screen, but, in order
that the screen be ``thin,'' it must surely be the case that its
thickness is $\Delta L \lesssim 1$~pc.

Suppose we require that the scattering contributed by such a clump not
make a significant contribution to the angular broadening.  For
illustration purposes, we adopt the amount of broadening contributed
by the clump to be $\theta_{s,\mathrm{cl}} \sim 0.2$~mas, which would
be only a 10\% contribution to the typical broadening diameter that we
measure.  The resulting scattering measure~SM \citep{cl02} is then
$\mathrm{SM}_{\mathrm{cl}} \sim 10^{-4.8}$~kpc~m${}^{-20/3}$.  If the
clump has $C_n^2 = 0.5$~m${}^{-20/3}$, the implied thickness is
0.05~pc ($\sim 10^4$~AU), which would certainly qualify as ``thin.''
Moreover, following \cite{rqwkw95} and \cite{r02}, it can be shown
that a more distant scattering clump tends to produce a lower
scintillation index.

Thus, our picture is one in which the Galactic disk contains (small)
``clumps'' of scattering material.  Lines of sight through the disk
are scattered by the broadly distributed ionized interstellar medium,
so that AGN over the range of latitudes that we observe have
measurable scattering diameters.  Some (many?) lines of sight pass
through one of these clumps, and AGN having compact enough components
are then observed to scintillate.  However, the clumps are small
enough that they produce effectively no additional broadening.  This
scenario is also broadly consistent with the notion of ``clumps'' of
material producing extreme scattering events \citep{fdjh87} and
parabolic arcs in pulsar dynamic spectra \citep{hsabeh05}.

We can also use the difference between the scintillating and
non-scintillating sources to set quantitative limits on the amount of
radio-wave scattering contributed by the \hbox{IGM}.  We adopt 0.5~mas
at~1~GHz ($\approx 3\sigma$ from Table~\ref{tab:stats}) as the upper
limit on the difference in the amount of scattering between the two
populations.  The implied scattering measure is $\mathrm{SM} \lesssim
10^{-4}$~kpc~m${}^{-20/3}$ \citep{cl02}.  In turn, the scattering
measure is given by
\begin{equation}
\mathrm{SM} = C_{\mathrm{SM}}\overline{Fn_e^2}D,
\label{eqn:sm}
\end{equation}
where $D$ is the distance, $F$ is a fluctuation parameter
encapsulating aspects of the microphysics of the plasma, $n_e$ is the
electron density, and $C_{\mathrm{SM}} = 1.8$~m${}^{-20/3}$~cm${}^6$
is a constant.  For a characteristic redshift of approximately unity
(Figure~\ref{fig:z}), the equivalent (angular-size) distance is $D
\approx 1.5$~Gpc, implying $\overline{Fn_e^2} \lesssim 10^{-10.5}$~cm${}^{-6}$.

For a baryonic matter density $\Omega_bh^2 = 0.127$ \citep{sbd+06},
the intergalactic electron density can be no larger than
$\overline{n_e} < 2.2 \times 10^{-7}$~cm${}^{-3}$, assuming that
helium is fully ionized \citep{sah02}.  Thus, we require $F \lesssim
10^3$, so as not to violate the inferred limits on scattering.  For
reference, in the diffuse Galactic \hbox{ISM}, $F \approx 0.2$, and in
the Galactic spiral arms, $F \sim 10$.  In turn, the $F$ parameter is
\begin{equation}
F = \zeta\epsilon^2\eta^{-1}\ell_0^{-2/3},
\label{eqn:f}
\end{equation}
where $\zeta$ is the normalized second moment of the fluctuations,
$\epsilon$ is the fractional variance in $n_e$ within the plasma,
$\eta$ is the filling factor, and $\ell_0$ is the largest scale on
which the density fluctuations occur (or outer scale, if the plasma is
turbulent), in parsec units.  Assuming that $\zeta \sim \epsilon \sim
1$, we conclude that $\eta\ell_0^{2/3} \gtrsim 10^{-3}$.

The IGM is thought to be permeated by shocks \citep{dco+01}, which
might be expected to drive $\eta \to 1$.  Given the larger scales
available in the \hbox{IGM}, $\ell_0 \sim 1$~Mpc would not be
unreasonable.  We are forced to conclude that the current limits on
intergalactic scattering, while broadly consistent with the current
picture of the \hbox{IGM}, do not yet place significant constraints on
its properties.

While we find no indications of intergalactic scattering, future
observations are warranted.  In particular, if a scintillating AGN
is found close to the line of sight to a pulsar, a comparison between
the two lines of sight would provide strong constraints on the amount
of Galactic vs.\ intergalactic scattering.  Also, higher-sensitivity
observations (e.g., with the very long baseline High Sensitivity
Array or HSA) targeting scintillating AGN with larger diameters may
provide additional constraints.  Many of the AGN with the largest
diameters are not detected at the lower frequencies, frequencies at
which the VLBA alone has a relatively low sensitivity.  The HSA could
be used to verify whether these AGN do indeed have such large
scattering diameters or assess to what extent intrinsic structure
contaminates the scattering diameter estimates.

We summarize our findings as follows.  In our sample of~58 \hbox{AGN},
approximately 75\% of the sample exhibit intraday variability
(interstellar scintillation) with the other 25\% not showing intraday
variability.  Interstellar scattering is measurable for most of these
\hbox{AGN}, and the typical broadening diameter is~2~mas.
Scintillating AGN are typically at lower Galactic latitudes than the
non-scintillating \hbox{AGN}, consistent with the scenario that
intraday variability is a propagation effect from the Galactic
interstellar medium.  The magnitude of the inferred interstellar
broadening measured toward the scintillating \hbox{AGN}, when scaled
to higher frequencies, is comparable to that determined from analyses
of the light curves for the more well-known intraday variable sources.
However, we find no difference in the amount of scattering measured
toward the scintillating versus non-scintillating \hbox{AGN}.  A
consistent picture is one in which the scintillation results from
localized regions (``clumps'') distributed throughout the Galactic
disk, but which individually make little contribution to the angular
broadening.  In our sample, 63\% of the AGN have measured redshifts.
At best, a marginal trend is found for scintillating
(non-scintillating) AGN to have smaller (larger) angular diameters at
higher redshifts.  Finally, while broadly consistent with the scenario
of a highly turbulent intergalactic medium, our observations do not
place significant constraints on its properties.

\acknowledgements

We thank the referee for suggestions that improved the presentation of
these results.
The National Radio Astronomy Observatory is a facility of the National
Science Foundation operated under cooperative agreement by Associated
Universities, Inc.  This research has made use of the United States
Naval Observatory (USNO) Radio Reference Frame Image Database (RRFID);
NASA's Astrophysics Data System bibliographic services; the SIMBAD
database, operated at \hbox{CDS}, Strasbourg, France; and the
NASA/IPAC Extragalactic Database (NED) which is operated by the Jet
Propulsion Laboratory, California Institute of Technology, under
contract with the National Aeronautics and Space Administration.
Basic research in radio astronomy at the NRL is supported by the NRL
Base funding.

\end{document}